\documentclass[12pt]{article}
\usepackage{amsmath}
\usepackage{graphicx,psfrag,epsf}
\usepackage{enumerate}
\usepackage{url}
\usepackage{graphicx}
\usepackage{hyperref}
\usepackage{float}
\usepackage[square,numbers]{natbib}
\usepackage{booktabs}

\newcommand{\blind}{0}

\def\spacingset#1{\renewcommand{\baselinestretch}%
{#1}\small\normalsize} \spacingset{1}

  \title{\bf XDC Staking and Tokenomics - Improvement Proposal: Enhancing Sustainability and Decentralization on the Eve of XDC 2.0}
  \author{
  Van Khanh Nguyen\\
  \texttt{13520392@ms.uit.edu.vn}
}
\begin{document}

  \maketitle

\if1\blind
{

  \begin{center}
\end{center}
  \medskip
} \fi

\bigskip
\begin{abstract}
As the XDC network celebrates five years of stable mainnet operation and prepares for the highly anticipated launch of XDC 2.0, this research proposes a comprehensive improvement plan for the network's staking and tokenomics mechanisms. Our analysis reveals opportunities to optimize the current model, ensuring a more sustainable, decentralized, and resilient ecosystem. We introduce novel concepts, including validator NFTs, decentralized governance, and utility-based tokenomics, to increase validator node liquidity and promote staking participation. Our proposal aims to establish a robust foundation for XDC 2.0, fostering a thriving ecosystem that rewards validators, stakeholders, and users alike. By addressing the intricacies of staking and tokenomics, this research paves the way for XDC to solidify its position as a leading decentralized network, poised for long-term success and growth.
\end{abstract}

\newpage
\spacingset{1.45}
\section{Introduction}
\label{sec:intro}

As the blockchain ecosystem continues to evolve, the importance of staking as a mechanism for securing and decentralizing Proof of Stake (PoS) networks cannot be overstated. Staking, which involves validators locking up their tokens or coins as collateral to participate in the validation process, has become a crucial component of many PoS-based protocols. However, as the XDC network prepares to launch its highly anticipated XDC 2.0 upgrade, it is essential to revisit the core concepts of staking and explore ways to enhance its staking mechanism to ensure long-term sustainability, security, and decentralization.

In this research, we will explore the fundamental principles of staking, examining its benefits, challenges, and opportunities for improvement. We will also draw inspiration from other prominent PoS-based projects, such as Ethereum, to identify best practices and innovative approaches that can be applied to the XDC network. By analyzing the staking models of these projects, we will gain valuable insights into how to optimize staking incentives, increase validator participation, and promote a healthier and more decentralized network. Through this exploration, we aim to provide a comprehensive roadmap for enhancing the staking mechanism of the XDC network, paving the way for a more robust, secure, and decentralized ecosystem that supports the growing demands of the blockchain industry.

\section{XDC Network staking analysis} 
\hypertarget{a.current-model}{%
\paragraph{a. Current model}\label{a.current-model}}

The consensus mechanism of the XDC network is designed to operate with
validator masternodes\cite{xdctokenomics} and standby masternodes. 

\begin{itemize}

    \item
      \textbf{Validator Masternodes:}
\begin{itemize}
    \item \textbf{Entities:} Operate as the primary network validators.
    \item \textbf{Income Sources:} Earnings are derived from a share of transaction fees and an annual income of 10\% based on their staked amount.
    \item \textbf{Income Formula:} Total Earnings = Share of txFee + (10\% x Staked XDC amount)
\end{itemize}
\end{itemize}

\begin{itemize}

\newpage

\item
  \textbf{Standby Masternodes:}

  \begin{itemize}
  
  \item
    \textbf{Entities:} Serve as backup validators and are activated when
    primary validators are unavailable.
  \item
    \textbf{Income Sources:} Receive a fixed income of 8\% annually
    based on their staked amount.
  \item
    \textbf{Income Formula:} Total Earnings = 8\% x Staked XDC amount
  \end{itemize}
\item
  To sustain the annual income distributions for both Validator and
  Standby Masternodes, the XDC network needs to emit funds annually
  calculated as follows:

  \begin{itemize}
  
  \item
    For Validator Masternodes: 108 x 10,000,000 x 10\%
  \item
    For Standby Masternodes: 108 x 10,000,000 x 8\%
  \item
    Total emission = (108 x 10,000,000 x 10\%) + (108 x 10,000,000 x
    8\%) = 194,400,000
  \item
    With the total XDC supply as of August 20, 2024 is 37,936,724,971,
    we can calculate the annual inflation of XDC now is as below
  \item
  Annual inflation rate = (Total emission / Total supply) x 100\% =
  (194,400,000/37,936,724,971) x 100\% = 0.512\%
\end{itemize}

The inflation rate of approximately 0.512\% for the XDC network is
relatively low compared to many other cryptocurrencies. This modest
inflation suggests a sustainable growth model for the network, aligning
with the goal of maintaining a stable increase in supply without
significantly diluting existing holders.

\hypertarget{b.-new-model}{%
\paragraph{b. New model}\label{b.-new-model}}

The proposed model introduces a significant change by increasing the
daily transaction fee to generate more revenue for validator
masternodes. This increase is predicated on the formula:

Daily transaction fee = Gas fee x Avg gas used x daily number of
transactions.

Under the new model, the daily transaction fee would increase by a
factor of 50. To achieve this, we could either increase the gas fee or
the daily number of transactions by the same factor. Among these
options, raising the gas fee is more feasible within the consensus
mechanism.

However, this approach has a substantial downside. Users of the XDC
network, who are accustomed to low transaction fees, would face a
50-fold increase in gas fees. Despite this significant rise, the annual
income for each active validator masternode would increase by less than
0.5\%, with no additional revenue for standby validators. This disparity
highlights a critical imbalance in the proposed economic model.
\section{General Concepts of Staking}

In a PoS network\cite{PoS}, staking involves validators locking up a certain
amount of their tokens or coins as collateral to participate in the
validation process. The validators are chosen to create a new block
based on the amount of tokens they have staked, also known as their
``stake.'' The validator with the highest stake has the highest chance
of being chosen to create a new block. Once a validator is chosen, they
create a new block and add it to the blockchain, and in return, they
receive a reward in the form of transaction fees and/or a block reward.

Staking has several benefits, including:

\begin{enumerate}
\def\labelenumi{\arabic{enumi}.}

\item
  \textbf{Energy Efficiency}: PoS networks are more energy-efficient
  compared to traditional Proof of Work (PoW) networks, which require
  massive amounts of energy to power their nodes.
\item
  \textbf{Faster Transaction Times}: PoS networks typically have faster
  transaction times compared to PoW networks, making them more suitable
  for everyday transactions.
\item
  \textbf{Increased Security}: The more validators that participate in
  the staking process, the more secure the network becomes, as it would
  require a significant amount of resources to launch a 51\% attack.
\end{enumerate}

\textbf{Categories of Staking}

Over time, different categories of staking have emerged, each with its
unique features and benefits.

\hypertarget{a.-traditional-staking}{%
\paragraph{a. Traditional Staking}\label{a.-traditional-staking}}

Traditional staking is the most common form of staking, where validators
lock up their tokens or coins in a wallet or a staking pool to
participate in the validation process. The validators are chosen to
create a new block based on their stake, and they receive a reward for
their participation.

\textbf{Example:} Tezos (XTZ) is a popular PoS blockchain that uses
traditional staking. Validators, also known as ``bakers,'' stake their
XTZ tokens to participate in the validation process and earn rewards in
the form of new XTZ tokens.

\hypertarget{b.-liquid-staking}{%
\paragraph{b. Liquid Staking}\label{b.-liquid-staking}}

Liquid staking is a newer concept that allows users to stake their
tokens or coins while still maintaining liquidity. In traditional
staking, users must lock up their tokens for a certain period, which can
range from a few days to several weeks or even months. Liquid staking,
on the other hand, allows users to stake their tokens while still being
able to use them for other purposes, such as trading or lending.

\textbf{Example:} Lido Finance is a liquid staking platform that allows
users to stake their Ethereum (ETH) tokens while still maintaining
liquidity. Users can stake their ETH tokens and receive a tokenized
staking derivative, which can be traded on decentralized exchanges
(DEXs) or used as collateral for loans.

\hypertarget{c.-defi-staking}{%
\paragraph{c. DeFi Staking}\label{c.-defi-staking}}

DeFi staking is a form of staking that combines the benefits of
decentralized finance (DeFi) with traditional staking. In DeFi staking,
users can stake their tokens or coins in decentralized lending
protocols, such as Compound or Aave, to earn interest on their deposits.
The interest earned is typically in the form of additional tokens or
coins, which can then be staked to earn even more rewards.

\textbf{Example:} Compound is a popular DeFi lending protocol that
allows users to stake their tokens, such as Ethereum (ETH) or DAI, to
earn interest on their deposits. Users can then stake their earned
interest to earn even more rewards, creating a compounding effect.

\hypertarget{d.-re-staking}{%
\paragraph{}{d.~Re-staking}}\label{d.-re-staking}

Re-staking is a form of staking that involves re-delegating staking
rewards to the original staking pool or another staking pool on the
network. This allows users to compound their staking rewards, earning
even more rewards on top of their initial stake.

\textbf{Example:} EigenLayer is a decentralized staking protocol that
allows users to re-stake their staking rewards to earn even more
rewards. Users can re-delegate their rewards to the original staking
pool or another staking pool on the network, creating a compounding
effect.

\section{Ethereum case study} Ethereum Improvement Proposal (EIP)
1559, introduced in August 2021 as part of Ethereum's London Hard Fork,
significantly altered how transaction fees are calculated and paid on
the Ethereum network. This proposal aimed to make Ethereum's fee market
more predictable and to improve the blockchain's overall economic
efficiency. Here's a detailed breakdown of its components and
implications:

\hypertarget{a.-dual-fee-structure}{%
\paragraph{a. Dual-Fee Structure}\label{a.-dual-fee-structure}}

EIP-1559\cite{ethereumresearch} introduces a dual-fee structure consisting of a ``base fee''
and an ``inclusion fee'' (also known as a tip or priority fee):

\begin{itemize}
\item
  \textbf{Base Fee:} This is a mandatory fee that is burned (removed
  from circulation), calculated algorithmically, and adjusted by the
  protocol based on network congestion. After every block, the base fee
  can increase or decrease by up to 12.5\% depending on whether the
  previous block was more or less than 50\% full.
\item
  \textbf{Inclusion Fee (Tip):} This is an optional fee paid to miners
  to prioritize a transaction over others. This fee is not burned and is
  given directly to the miners for including the transaction in a block.
\end{itemize}

\hypertarget{b.-fee-burning}{%
\paragraph{b. Fee Burning}\label{b.-fee-burning}}

The base fee is burned, meaning it is permanently removed from the
Ethereum supply. This mechanism serves multiple purposes:

\begin{itemize}
\item
  \textbf{Reduces Ethereum Inflation:} By burning the base fee, EIP-1559
  introduces a deflationary mechanism to Ethereum's economy, potentially
  increasing the value of Ether over time.
\item
  \textbf{Mitigates Economic Abstraction:} By requiring the base fee to
  be paid in Ether (ETH), it ensures that ETH remains central to the
  Ethereum network, as opposed to allowing transactions to be paid with
  other tokens.
\end{itemize}

\hypertarget{c.-block-size-variability}{%
\paragraph{c. Block Size Variability}\label{c.block-size-variability}}

Under EIP-1559, the concept of a fixed block size is replaced with a
more flexible model:

\begin{itemize}

\item
  \textbf{Target Block Size and Elasticity:} The protocol targets a
  block size (e.g., 15 million gas) but allows blocks to expand up to
  twice the target size (e.g., 30 million gas) to accommodate sudden
  spikes in network demand. This flexibility helps in managing
  congestion and reduces the variability in transaction fees during peak
  times.
\end{itemize}

\textbf{Reward and Penalty model} With the transition to PoS, the
issuance mechanism shifted from mining to staking. Validators now secure
the network by staking ETH, and issuance is tied to these staking
activities.

\begin{itemize}

\item
  \textbf{Per-Block Issuance}: The rate at which new tokens are issued
  per block varies based on the total volume of ETH staked across the
  network. Validators receive these new tokens as a reward for their
  participation and contribution to network security. This dynamic
  issuance rate ensures that the incentive for validators adjusts with
  changes in network participation.
\item
  \textbf{Validator Tip}: This fee incentivizes validators to prioritize
  the inclusion of a user's transaction in a block. Unlike the base fee,
  which is algorithmically determined and burned, the validator tip is
  not subject to burning and goes directly to the validators as a reward
  for processing and validating transactions. This mechanism ensures
  that validators are compensated for their computational efforts,
  especially during periods of high network congestion, when users might
  be willing to pay a premium for faster transaction processing.
\item
  \textbf{Penalties and Slashing}: Validators can lose a portion of
  their stake for actions that could harm the network, such as being
  offline (inactivity leak) or attempting to corrupt the network
  (slashing).
\end{itemize}

Therefore, the total validator rewards is as follows: \textbf{Total
rewards} = \emph{Per-Block Issuance} + \emph{Validator Tip} -
\emph{Penalties and Slashing}

The annual net inflation of the network is as follows: \textbf{Net ETH
supply growth} = \emph{Total staked ETH} x \emph{issuance rate} -
\emph{Total base fee}

The current inflation rate of ETH could be seen in this link
https://ultrasound.money/

\section{Other case studies\cite{staking} }
\hypertarget{a.Traditional staking}{%
\paragraph{a. Traditional staking}\label{a.traditional-staking}}
\begin{itemize}

\item
  \textbf{AION}

  \begin{itemize}

  \item
    The individual AION rewards depends on the Block Reward, Block Time,
    Daily Network Rewards and Total Staked. Every block one validator is
    randomly selected to create a block, whereas 1 staked or delegated
    token counts as one ``lottery ticket''. The selected validator has
    the right to create a new block and broadcast them to the network.
    The Validator then receives the 50\% of the block reward and the
    fees of all transactions (network rewards) successfully included in
    this block, whereas the PoW Miner receives the other 50\%.
  \end{itemize}
\item
  \textbf{Algorand}

  \begin{itemize}

  \item
    Rewards in the form of algos are granted to Algorand users for a
    variety of purposes. Initially, for every block that is minted,
    every user in Algorand receives an amount of rewards proportional to
    their stake in order to establish a large user base and distribute
    stake among many parties. As the network evolves, the Algorand
    Foundation will introduce additional rewards in order to promote
    behavior that strengthens the network, such as running nodes and
    proposing blocks.
  \end{itemize}
\item
  \textbf{BitBay}

  \begin{itemize}

  \item
    The individual BitBay rewards depends on the Block Reward, Block
    Time, Daily Network Rewards and Total Staked. Every block is
    randomly selected whereas 1 staked coin counts as one ``lottery
    ticket''. The selected staker has the right to create a new block
    and broadcast it to the network. He then receives the block reward
    and the fees of all transactions successfully included in this
    block.
  \end{itemize}
\item
  \textbf{Dash}

  \begin{itemize}

  \item
    Dash blockchain consensus is achieved via Proof of Work +
    Masternodes. Investors can leverage their crypto via operating
    masternodes. Miners are rewarded for securing the blockchain and
    masternodes are rewarded for validating, storing and serving the
    blockchain to users.
  \end{itemize}
\item
  \textbf{EOS}

  \begin{itemize}

  \item
    EOS has a fixed 5\% annual inflation. 4\% goes to a savings fund,
    which might distribute the funds to the community later on. 1\% goes
    to Block producers and Standby Block Producers. Out of the 1\% that
    are given to block producers, only 0.25\% will go to the actual 21
    producers of the blocks. The other 0.75\% will be shared among all
    block producers and standby block producers based on how many votes
    they receive and with a minimum of 100 EOS/day.
  \end{itemize}
\item
  \textbf{Fantom}

  \begin{itemize}

  \item
    The individual reward of staking fantom depends on the Total Staked
    ratio. Transactions are packaged into event blocks. In order for
    event blocks to achieve finality, event blocks are passed between
    validator nodes that represent at least 2/3rds of the total
    validating power of the network. A validator's total validating
    power is primarily determined by the number of tokens staked and
    delegated to it. A validator earns rewards each epoch for each event
    block signed according to it's validating power. By delegating,
    investors can increase the share of their validator proportionally
    to the balance of their account. They will receive rewards
    accordingly and share them with investors after taking the
    commission.
  \end{itemize}
\item
  \textbf{Livepeer}

  \begin{itemize}

  \item
    Every livepeer (LPT) token holder has the right to delegate their
    tokens to an Orchestrator node for the right to receive both
    inflationary rewards in LPT and fees denominated in ETH from work
    completed by that node. The individual LTO rewards depends on the
    Network Rewards (Transaction Fees spent on the Network) and the
    Total Staked. Every block one staking node operator is randomly
    selected to create a new block, whereas 1 staked token counts as one
    ``lottery ticket''. The staker receives the fees of all transactions
    successfully included in this block. Staking Node Operators share
    the rewards with their delegators after deducting a commission.
  \end{itemize}
\item
  \textbf{NEM}

  \begin{itemize}

  \item
    NEM blockchain consensus is achieved via Proof of Importance.
    Investors can leverage their crypto via harvesting. To harvest NEM
    coins it is recommended to run the official NEM Core wallet with an
    entire copy of the blockchain on the stakers' computer or a Virtual
    Private Server (VPS). The individual NEM harvesting rewards depends
    on the Daily Network Rewards and Total Staked. For every block, the
    staker is randomly selected whereas 1 staked coin counts as one
    ``lottery ticket''. The selected staker has the right to create a
    new block and broadcast it to the network. The staker then receives
    the fees of all transactions successfully included in this block.
  \end{itemize}
\item
  \textbf{NEO}

  \begin{itemize}

  \item
    Everyone who holds NEO will automatically be rewarded by GAS. GAS is
    produced with each new block. In the first year, each new block
    generates 8 GAS, and then decreases every year until each block
    generates 1 GAS. This generation mechanism will be maintained until
    the total amount of GAS reaches 100 million and no new GAS will be
    generated.
  \end{itemize}
\item
  \textbf{Nuls}

  \begin{itemize}

  \item
    Nuls blockchain consensus is achieved via Proof of Stake +
    Masternodes. Investors can leverage their crypto via staking. The
    amount earned is variable based on the current blockchain metrics
    like the amount of stakers (Total Staked ratio). Investors can stake
    Nuls into a project's nodes and earn their token as a reward, while
    the project earns Nuls as a reward. Some projects offer to stake
    with just 5 Nuls as the minimum.
  \end{itemize}
\item
  \textbf{Polkadot}

  \begin{itemize}

  \item
    Delegators in Polkadot are called Nominators. Anyone can nominate up
    to 16 validators, who share rewards if they are elected into the
    active validators set. The process is a single-click operation
    inside the wallet. The current reward rate for validators is
    determined by the current Total Staked ratio. The less DOT is being
    staked, the higher are the rewards.
  \end{itemize}
\item
  \textbf{Qtum}

  \begin{itemize}

  \item
    Qtum blockchain consensus is achieved via Proof of Stake 3.0. The
    individual reward depends on the Block Reward, Block Time, Daily
    Network Rewards and Total Staked. Every block is randomly selected
    whereas 1 staked coin counts as one ``lottery ticket''. The selected
    staker has the right to create a new block and broadcast it to the
    network. The staker then receives the block reward and the fees of
    all transactions successfully included in this block.
  \end{itemize}
\item
  \textbf{Synthetix Network}

  \begin{itemize}

  \item
    Synthetix Network Token blockchain consensus is achieved via the
    Ethereum Blockchain. Investors can leverage their crypto via
    staking. SNX holders can lock their SNX as collateral to stake the
    system. Synths are minted into the market against the value of the
    locked SNX, where they can be used for a variety of purposes
    including trading and remittance. All Synth trades on Synthetix
    Exchange generate fees that are distributed to SNX holders,
    rewarding them for staking the system.
  \end{itemize}
\item
  \textbf{Tezos}

  \begin{itemize}

  \item
    Tezos blockchain consensus is achieved via Liquid Proof of Stake.
    Investors can leverage their crypto via baking or delegating. There
    are a number of tokens that use a similar mechanism, including
    iotex, irisnet, etc.
  \end{itemize}
\item
  \textbf{Tron}

  \begin{itemize}

  \item
    Tron reward depends on the Block Rewards, Endorsement Rewards, Block
    Time, Daily Network Rewards and Total Staked. Every block is
    randomly selected to bake a block and 32 stakers are selected to
    endorse a block, whereas 1 staked coin counts as one ``lottery
    ticket''. The selected stakers have the right to create or endorse
    new block and broadcast them network. The Baker then receives the
    block reward and the fees of all transactions successfully included
    in this block. The Endorsers receive the endorsement rewards.
  \end{itemize}
\item
  \textbf{Wanchain}

  \begin{itemize}

  \item
    Wanchain blockchain consensus is achieved via Galaxy Proof-of-Stake.
    The individual WAN rewards depends on the Foundation Rewards, Daily
    Network Rewards and Total Staked. At the beginning of each protocol
    cycle (epoch), two groups, the RNP (Random Number Proposer) group
    and the EL (Epoch Leader) group, are selected from all validators. 1
    staked or delegated token counts as one ``lottery ticket'' to be
    selected. The two groups equally share the Foundation Rewards and
    Transaction Fees (Network Rewards). The Foundation Rewards consists
    of 10\% of the outstanding Wanchain Token Supply and are decreasing
    by 13.6\% each year, whereas the Network Rewards are expected to
    rise alongside wider network usage.
  \end{itemize}
\end{itemize}

\hypertarget{b.-liquid-staking-1}{%
\paragraph{b. Liquid staking}\label{b.-liquid-staking-1}}

\begin{itemize}
\item
  \textbf{Lido}

  \begin{itemize}
  
  \item
    Offers decentralized Ethereum staking.
  \item
    Users can stake any amount of ETH without the need to run their own
    infrastructure or meet a minimum deposit.
  \item
    Lido issues stETH to represent the staked ETH, allowing users to
    retain liquidity and engage in other DeFi activities.
  \item
  \end{itemize}
\item
  \textbf{Rocket Pool}:

  \begin{itemize}
  
  \item
    Offers decentralized Ethereum staking.
  \item
    Users can run their own node or stake ETH with a minimum of 16 ETH
    for node operators and 0.01 ETH for regular stakers.
  \item
    It provides rETH as a staking derivative, similar to Lido's stETH.
  \end{itemize}
\item
  \textbf{Ankr}:

  \begin{itemize}
  
  \item
    Provides staking services for multiple chains, including Ethereum.
  \item
    Offers aETH as a staking derivative, enabling liquidity for stakers.
  \item
    Focuses on providing infrastructure solutions as well as staking.
  \end{itemize}
\item
  \textbf{StakeWise}:

  \begin{itemize}
  
  \item
    Another decentralized option for Ethereum staking.
  \item
    Offers pooled staking with no minimum deposit, similar to Lido.
  \item
    Stakers receive sETH2, which represents their staked ETH and accrued
    rewards.
  \end{itemize}
\item
  \textbf{Frax Finance (Frax Ether)}:

  \begin{itemize}
  
  \item
    Introduced Frax Ether, a liquid staking derivative of Ethereum.
  \item
    Emphasizes a decentralized and algorithmic approach to staking and
    derivatives.
  \item
    Integrates with the broader Frax ecosystem for additional
    functionalities.
  \end{itemize}
\item
  \textbf{Marinade Finance (Solana-focused)}:

  \begin{itemize}
  
  \item
    While it's primarily for Solana, it offers a model similar to Lido
    for SOL staking.
  \item
    Provides mSOL as a liquid staking token to represent staked SOL.
  \item
    Aims to decentralize the staking process and enhance liquidity.
    \#\#\#\# c.~DeFi staking
  \end{itemize}
\item
  \textbf{Aave}:

  \begin{itemize}
  
  \item
    Aave is a decentralized lending protocol that allows users to lend
    and borrow a variety of cryptocurrencies.
  \item
    Users can deposit DAI to earn interest, which is determined by
    supply and demand dynamics in the market.
  \end{itemize}
\item
  \textbf{Compound}:

  \begin{itemize}
  
  \item
    Compound operates similarly to Aave, providing a decentralized
    finance platform for lending and borrowing.
  \item
    Users can supply DAI to the Compound pool and earn interest, as well
    as cDAI, which represents the staked DAI and accrued interest.
  \end{itemize}
\item
  \textbf{Yearn.finance}:

  \begin{itemize}
  
  \item
    Yearn.finance optimizes the yield earning capabilities of the assets
    deposited.
  \item
    It automatically moves DAI between different lending services like
    Aave, Compound, and others to maximize the interest earnings.
  \end{itemize}
\item
  \textbf{MakerDAO}:

  \begin{itemize}
  
  \item
    While primarily known for creating DAI, MakerDAO also offers
    opportunities to engage DAI in its ecosystem to earn savings through
    the DAI Savings Rate (DSR).
  \item
    Users can lock their DAI in the DSR to earn a variable interest rate
    directly through the MakerDAO protocol.
  \end{itemize}
\item
  \textbf{Curve Finance}:

  \begin{itemize}
  
  \item
    Curve is a decentralized exchange for stablecoins that uses an
    automated market maker (AMM) to manage liquidity.
  \item
    Users can deposit DAI into Curve liquidity pools to facilitate
    trading and earn trading fees, in addition to potential CRV token
    rewards. Certainly! Based on the information provided about Fathom
    Protocol, here's a simplified summary similar in format to the
    previous examples:
  \end{itemize}
\item
  \textbf{Fathom Protocol:}

  \begin{itemize}
  
  \item
    Offers liquidity solutions for both retail and institutional
    participants, focusing on integrating real-world assets (RWA) with
    cryptocurrency financing.
  \item
    Users can deposit XDC and staked XDC to mint FXD, a USD-stablecoin.
    This process allows users to utilize their XDC holdings to generate
    stablecoin liquidity.
  \item
    FXD can be used to participate in yield-generating activities
    through RWA Vaults and Trade Finance Pools. Additionally, crypto and
    RWA holders can borrow FXD using their assets as collateral,
    facilitating flexible financial operations. \#\#\#\#
    \textbf{d.~Re-staking} \#\#\#\# EigenLayer is a cutting-edge
    protocol that leverages existing staked assets on Ethereum to
    provide additional security and functionality without requiring
    users to unstake their ETH. Here's a quick overview:
  \end{itemize}
\item
  \textbf{EigenLayer:}

  \begin{itemize}
  \item
    Allows users to re-stake their already staked Ethereum to support
    additional protocols on top of the Ethereum base layer, enhancing
    network security and computational capabilities without compromising
    their initial staking position.
  \item
    Participants can contribute to the security of multiple protocols
    simultaneously, increasing the utility of their staked ETH and
    opening up new avenues for earning rewards.
  \item
    By allowing staked ETH to be leveraged for multiple purposes,
    EigenLayer introduces a novel approach to maximize the efficiency
    and utility of staked assets within the Ethereum ecosystem.
  \end{itemize}
\end{itemize}

\section{Some reward schemes for XDC staking models}

\begin{itemize}

\item
  \textbf{Reward scheme 1}:

  \begin{itemize}
  \item
    Reward both validator masternodes and standby masternodes flat rate
    9\% staked XDC amount annually. Validator masternodes will get
    additional rewards for the blocks they validate. In that case, the
    income for each type of validator would be as below

\begin{verbatim}
Validator masternodes earnings  = Share of txFee + (9% x Staked XDC amount)
Standby masternodes earnings  = 9% x Staked XDC amount
\end{verbatim}
  \item
    This reward scheme is predicated on the requirement that both active
    validator masternodes and standby masternodes stake an equal amount
    and maintain identical levels of availability to ensure continuous
    transaction validation under any circumstances. This approach is
    designed to incentivize a greater number of nodes to become standby
    validators, enhancing network resilience and reliability.
  \end{itemize}
\item
  \textbf{Reward scheme 2}:

  \begin{itemize}
  \item
    Another reward scheme permits users to tip validator masternodes,
    which allows their transactions to be validated more quickly
    relative to others. Under this scheme, the transaction fee is burnt
    to prevent inflation. Standby validators are not left out; they
    continue to earn an annual 8\% return on their staked amount,
    ensuring consistent rewards regardless of active transaction
    validation. In that case, the income for each type of validator
    would be as below

\begin{verbatim}
Validator masternodes earnings  = Validator tip + (10% x Staked XDC amount)
Standby masternodes earnings  = 8% x Staked XDC amount
\end{verbatim}
  \item
    This reward scheme effectively mitigates network inflation while
    ensuring that both Validator and Standby masternodes receive
    reasonable compensation based on their staked amounts. However, for
    this scheme to function optimally, it necessitates the creation of
    high demand for rapid transactions within the network.
  \end{itemize}

\end{itemize}
\end{itemize}
\section{Some proposals for
tokenomics}

\textbf{a. Introducing Loyalty Factor for Node Validators}

\begin{itemize}

\item
  \textbf{Target:} Recruit more investors/node operators who trust and
  contribute to the growth of the XDC network.
\item
  \textbf{Detail:}

  \begin{itemize}
  
  \item
    Node operators can choose to stake 10,000,000 XDC to the network for
    a specified duration to earn a corresponding loyalty factor.
  \item
    The loyalty factor will be calculated based on the staking duration,
    with a maximum loyalty factor of 10 for a 10-year staking period.
  \item
    The loyalty factor will determine the validator's share of the total
    reward pool, ensuring that long-term contributors are incentivized
    and rewarded fairly.
  \item
    The pool of rewards for all validators will be 194,400,000 XDC, and
    each validator's reward will be calculated based on their loyalty
    factor.
  \end{itemize}
\end{itemize}

\textbf{Loyalty Factor Table:}

\begin{table}[ht]
\centering
\caption{Staking Duration and Loyalty Factors}
\label{tab:loyalty_factors}
\begin{tabular}{ll}
\toprule  
Staking Duration & Loyalty Factor \\
\midrule  
1 year  & 1  \\
2 years & 2  \\
3 years & 3  \\
5 years & 5  \\
10 years & 10 \\
\bottomrule  
\end{tabular}
\end{table}

\textbf{b. Withdrawal Fee Based on Withdrawal Time}

\begin{itemize}

\item
  \textbf{Target:} Reduce the real inflation rate of the network in the
  short term.
\item
  \textbf{Detail:}

  \begin{itemize}
  
  \item
    Instead of rewarding validators after each block, rewards will be
    sent to a smart contract and can be withdrawn after reaching a
    certain threshold.
  \item
    A withdrawal fee will be applied, which will be burnt, to discourage
    frequent withdrawals and reduce inflation.
  \item
    The withdrawal fee will be calculated based on the withdrawal amount
    and time, with lower fees for larger and longer-term withdrawals.
  \end{itemize}
\end{itemize}

\textbf{Withdrawal Fee Table:}

\begin{table}[ht]
\centering
\caption{Withdrawal Amounts and Fees}
\label{tab:withdrawal_fees}
\begin{tabular}{@{}ll@{}}
\toprule  
Withdrawal Amount & Withdrawal Fee \\
\midrule  
1,000 XDC    & 50\% \\
10,000 XDC   & 30\% \\
100,000 XDC  & 20\% \\
1,000,000 XDC & 0\% (after 1 year) \\
\bottomrule  
\end{tabular}
\end{table}

\textbf{c.~Reward Reinvestment Program}

\begin{itemize}

\item
  \textbf{Target:} Allow node validators to contribute to the growth of
  the ecosystem by introducing an investment program.
\item
  \textbf{Detail:}

  \begin{itemize}
  
  \item
    Node validators can invest their rewards in an ecosystem treasury at
    any time.
  \item
    The investment will be subject to a withdrawal fee, which will be
    contributed to the ecosystem treasury.
  \item
    The treasury will distribute funds to projects that generate more
    customers and transaction fees, based on their performance.
  \item
    Validators will receive a pool share for their contribution, which
    can be redeemed to earn rewards.
  \item
    The fee earned by each project from customer fees will be returned
    to the ecosystem treasury, allowing validators to redeem their pool
    share and earn rewards.
  \end{itemize}
\end{itemize}

\textbf{Reward Reinvestment Program Flow:}

\begin{enumerate}
\def\labelenumi{\arabic{enumi}.}

\item
  Node validators invest their rewards in the ecosystem treasury.
\item
  The investment is subject to a withdrawal fee, which is contributed to
  the treasury.
\item
  The treasury distributes funds to projects based on their performance.
\item
  Validators receive a pool share for their contribution.
\item
  Validators can redeem their pool share to earn rewards.
\item
  Project fees are returned to the treasury, allowing validators to
  redeem their pool share and earn rewards.
\end{enumerate}

Here is the enhanced proposal:

\textbf{d.~Dynamic Fee and Tip Mechanism}

\begin{itemize}

\item
  \textbf{Target:} Adjust the transaction fee dynamically based on
  network demand and performance to ensure a sustainable and scalable
  ecosystem.
\item
  \textbf{Detail:}

  \begin{itemize}
  
  \item
    Implement a dynamic fee mechanism that adjusts the transaction fee
    based on the network's transaction per second (TPS) rate.
  \item
    The fee will be calculated using a tiered system, with higher fees
    for higher TPS rates.
  \item
    The current average fee of 0.0002 XDC will be increased by a factor
    of 10 when the TPS rate reaches 5-12 times the current rate.
  \item
    When the TPS rate reaches 100 times the current rate, a tip
    mechanism will be introduced, allowing users to incentivize
    validators with additional fees for priority transaction processing.
  \end{itemize}
\end{itemize}

\textbf{Dynamic Fee Tier System:}

\begin{table}[ht]
\centering
\caption{TPS Rate and Fee Multiplier}
\label{tab:tps_fee}
\begin{tabular}{@{}ll@{}}
\toprule
TPS Rate & Fee Multiplier \\
\midrule
1-5       & 1x \\
5-12      & 10x \\
12-25     & 20x \\
25-50     & 50x \\
50-100    & 100x \\
100+      & Tip Mechanism \\
\bottomrule
\end{tabular}
\end{table}

\textbf{Tip Mechanism:}

\begin{itemize}

\item
  \textbf{Tip Rate:} 0.001-0.01 XDC per transaction (configurable)
\item
  \textbf{Tip Priority:} Transactions with higher tips will be
  prioritized for processing, ensuring faster confirmation times for
  critical transactions.
\end{itemize}

\textbf{Benefits:}

\begin{itemize}

\item
  \textbf{Increased Revenue for Validators:} Dynamic fees and tips will
  provide a new revenue stream for validators, incentivizing them to
  contribute to the network's growth and security.
\item
  \textbf{Scalability:} The dynamic fee mechanism will help manage
  network congestion, ensuring that the network can handle increasing
  transaction volumes without compromising performance.
\item
  \textbf{User Experience:} The tip mechanism will provide users with a
  way to prioritize their transactions, ensuring faster confirmation
  times for critical transactions.
\end{itemize}

By implementing a dynamic fee and tip mechanism, the XDC network can
adapt to changing demand and ensure a sustainable and scalable ecosystem
for all participants.

These enhanced proposals aim to incentivize node validators to
contribute to the growth of the XDC network, reduce inflation, and
promote ecosystem development.

\bibliography{references.bib}

\begin{thebibliography}{1}

\bibitem{xdctokenomics}
XDC Community.
\newblock Tokenomics.
\newblock \url{https://docs.xdc.community/get-started/xdc-design/tokenomics}, 2023.

\bibitem{ethereumresearch}
Kose John.
\newblock Economics of ethereum.
\newblock \url{https://papers.ssrn.com/sol3/papers.cfm?abstract_id=4783695}, 2024.

\bibitem{staking}
Ke~Tang Lin William~Cong, Zhiheng~He.
\newblock Staking, token pricing, and crypto carry.
\newblock \url{https://www.researchgate.net/publication/359796628_Staking_Token_Pricing_and_Crypto_Carry}, 2022.

\bibitem{PoS}
Fahad Saleh.
\newblock Blockchain without waste: Proof-of-stake.
\newblock \url{https://www.researchgate.net/publication/344935107_Blockchain_without_Waste_Proof-of-Stake}, 2020.

\end{thebibliography}

\end{document}